%
%
\magnification=\magstep1
\baselineskip=11pt plus .1pt minus .1pt
\hsize=12.5truecm
\vsize=19.0truecm  
\hfuzz=5pt\vfuzz=5pt
\tolerance=1000
\overfullrule=0pt
\parskip=0pt
\abovedisplayskip=3 mm plus6pt minus 4pt
\belowdisplayskip=3 mm plus6pt minus 4pt
\abovedisplayshortskip=0mm plus6pt minus 2pt
\belowdisplayshortskip=2 mm plus4pt minus 4pt
\predisplaypenalty=0
\clubpenalty=10000
\widowpenalty=10000
\parindent=2em
%
%
\font\pgnumfont=cmr9
\font\headlinefont=cmti9
\font\titlefont=cmbx10
\font\authorfont=cmr10
\font\addressfont=cmti9

\font\sumfont=cmr9

\font\rml=cmr9

\font\secfont=cmr10
\font\subsecfont=cmti10
\font\subsubsecfont=cmr10
\font\figfont=cmr9
\font\figheadfont=cmbx9
\font\tabfont=cmr9
\font\tabheadfont=cmbx9
\font\mainfont=cmr10
\font\petitrm=cmr9

%
%
%
\newtoks\TITLE \newtoks\AUTHOR \newtoks\ADDRESS \newtoks\SUMMARY
\newdimen\sumindent \sumindent=\parindent
\newtoks\KEYWORDS \newtoks\SUBMITTED \newtoks\ACCEPTED
\newtoks\SENDOFF
%

%
%
\newtoks\firstpage
\let\firstpage=Y
\newtoks\AUTHORHEAD \newtoks\ARTHEAD \newtoks\VOLUME \newtoks\PAGES
\if!\the\AUTHORHEAD!\AUTHORHEAD={\the\AUTHOR}\fi
\if!\the\ARTHEAD!\ARTHEAD={\the\TITLE}\fi
\footline={\hfil}
\headline={\ifodd\pageno\rightheadline \else\leftheadline\fi}
\def\leftheadline{\if Y\firstpage\firsthead\global\let\firstpage=N
  \else\lefthead\fi}
\def\rightheadline{\if Y\firstpage\firsthead\global\let\firstpage=N
  \else\righthead\fi}
\def\lefthead{\pgnumfont\number\pageno\hfil\headlinefont\the\AUTHORHEAD}
\def\righthead{\headlinefont\the\ARTHEAD\hfil\pgnumfont\number\pageno}
\def\firsthead{\headlinefont Baltic Astronomy,~vol.\the\VOLUME,
\the\PAGES,~\the\year .\hfil}
\voffset=2\baselineskip 
%

\newdimen\oldbaselineskip \oldbaselineskip=\baselineskip
\def\test#1{\newlinechar=`@\if!\the#1! \message{#1 not given@}\fi}%
\def\printheader{
  \parindent=0pt
  \null\vskip1.cm
  \test{\TITLE}
  \vbox{\baselineskip=15pt
    \titlefont\the\TITLE
    }
  \vskip8mm plus8mm
  \test{\AUTHOR}
  \authorfont\the\AUTHOR
  \vskip2mm
  \test{\ADDRESS}
  \addressfont\the\ADDRESS
  \vskip2mm
  \sumfont
  \if!\the\SENDOFF!\else\footnote{}{
 \the\SENDOFF}\fi
  \parindent=2em
  }
%
%
\newdimen\uppergap \newdimen\lowergap
\uppergap=5mm \lowergap=3mm
\newdimen\secind \newdimen\subsecind \newdimen\subsubsecind
\setbox0=\hbox{\secfont 9. }\secind=\wd0
\setbox0=\hbox{\subsecfont 9.9. }\subsecind=\wd0
\setbox0=\hbox{\subsubsecfont 9.9.9. }\subsubsecind=\wd0
\def\section#1{\goodbreak\par\vskip\uppergap
  \noindent\hangindent\secind\hangafter=1\secfont#1
  \vskip\lowergap\mainfont\par\nobreak}
\def\subsection#1{\goodbreak\par\vskip\uppergap
  \noindent\hangindent\subsecind\hangafter=1\subsecfont#1
  \vskip\lowergap\mainfont\par\nobreak}
\def\subsubsection#1{\goodbreak\par\vskip\uppergap
  \noindent\hangindent\subsubsecind\hangafter=1\subsubsecfont#1
  \vskip\lowergap\mainfont\par\nobreak}
%
%
%

%

%
\newdimen\tabind
\setbox0=\hbox{\tabheadfont Table 55.} \tabind=\wd0
\def\Table#1#2{\noindent
  \hangindent\tabind\hangafter=1
  \tabheadfont Table~#1.\tabfont #2
 \par
  \mainfont
  }
%
%
\def\References{\vskip\uppergap
\line{\secfont REFERENCES\hfill}
  \vskip0.8\lowergap
 \petitrm
  }
\def\ref{\goodbreak
\hangindent12pt\hangafter=1
\noindent\ignorespaces}
\def\endref{\egroup}

\def\ref{\goodbreak
\hangindent12pt\hangafter=1
\noindent\ignorespaces}
\def\endref{\egroup}
%
%
\def\byebye{\egroup\par\vfill\supereject\end}
%
%

%
%

\def\degr{\hbox{$^\circ$}}

\def\utw{\smash{\rlap{\lower5pt\hbox{$\sim$}}}}
\def\udtw{\smash{\rlap{\lower6pt\hbox{$\approx$}}}}

\def\sarcsec{\hbox{$^{\scriptscriptstyle\prime\prime}$}}

\newdimen\free\newdimen\shift
\def\Entry#1#2#3{\par\goodbreak\smallskip%
  \setbox1=\vbox{\advance\hsize by-10mm\parindent=0pt
    \def\\{\par}%
    \it#1. \rm#2}
  \line{\box1\hfill#3}\smallskip
}%
\newdimen\savesize

\def\shiftfigure #1#2#3#4#5{
    \vbox to #2 { \ifodd #5 \rightskip#4 \else\leftskip#4 \fi
                  \null\vfil
                  \figheadfont Fig.~#1.\figfont #3
                  \medskip
                }
                          }
\def\continued#1{
	\vfil\eject\null\vfil
	\line{\hfil\rml Table~#1~(continued)~~~}
		}
\year1997

%
%
\def\Straizys{Strai\v zys}

\def\Zdanavicius{Zdanavi\v cius}

%
\def\sm#1{\raise.3ex\hbox{$\scriptstyle #1$}}        
\def\ssm#1{\raise.3ex\hbox{$\scriptscriptstyle #1$}} 
\def\m{\ssm{-}} 
\def\ts{\thinspace}   
\def\ns{\enspace}     
\def\ms{\kern 1.0em}  
\def\nts{\kern -0.08em} 
\def\nns{\kern -0.15em} 
\def\ssd{\rlap{:}\ }  
\def\lsd{\rlap{::}\ } 
\def\Notes{\rlap{*}}  
\def\Ruler{\noalign{\smallskip\hrule\vskip2truemm}} 
\def\HeaderTableOneE{Star &$\scriptstyle\alpha$(2000)\ &$\scriptstyle\delta$(2000)\ 
&{\itn \ V\ } &{\itn U--P} &{\itn P--X} &{\itn X--Y} &{\itn Y--Z} &{\itn Z--V} &{\itn V--S} 
&{\itn n} &Sp\hfil\cr} 
\def\Hr{${}^{\rm \nts h}$\nns}  
\def\Min{${}^{\rm \nts m}$\nts} 
\def\Sec{${}^{\rm s}$}          
\def\Deg{$^{\rm o}$}            
\def\BlankLineTableOne{ & & & & & & & & & & & \cr }     
\def\HeaderTableTwo{{\itn V} &{\itn U--P} &{\itn P--X} &{\itn X--Y} &{\itn Y--Z} &{\itn Z--V} &{\itn V--S} \cr}
%
\def\stdev{
$$
\varepsilon=
{ 
	{\sigma}
	\over{
	\sqrt{ 
		{\raise.5ex\hbox{$ \textstyle \sum\limits^{\ } \displaystyle n $}}
		\over
		{\lower.5ex\hbox{\tenit N}} 
	}
	}
}
$$
}
%

%
%

\def\totalnumberofstars{96}
\def\numberofstars{60}
\def\ts{\thinspace}
\font\itn=cmti7
\font\rmn=cmr7
\year 1997
\VOLUME={~6}
\PAGES={1-6}
\pageno=1

\TITLE={THE SOUTHERN VILNIUS PHOTOMETRIC SYSTEM.
IV. THE E REGION STANDARD STARS}

\ARTHEAD={The Southern Vilnius Photometric System. IV.}

\AUTHOR={M.\ts C.\ts Forbes$^1$, R.\ts J.\ts Dodd$^2$ and D.\ts J.\ts
Sullivan$^1$}

\AUTHORHEAD={M.\ts C.\ts Forbes, R.\ts J.\ts Dodd and D.\ts J.\ts
Sullivan}

\ADDRESS={$^1$Victoria University of Wellington, P.O.Box 600,
Wellington, New~Zealand\break
$^2$Carter Observatory, P.O. Box 2909, Wellington, New Zealand}

\SUMMARY={This paper is the fourth in a series on the extension of the
Vilnius photometric system to the southern hemisphere.  Observations
were made of 60 stars in the Harvard Standard E regions to increase a
set of standard stars.}

\KEYWORDS={stars: multicolor photometry, standards -- methods:
observational -- techniques: photometric}

\SUBMITTED={July 21, 1996}

\printheader

\section{1. INTRODUCTION}

In the paper by Forbes, Dodd \& Sullivan (1993), hereafter referred to
as Paper~I, measurements of the equatorial stars, already observed in
the northern hemisphere Vilnius system, were made using Carter
Observatory's set of filters.  It was concluded that these could be
accurately transformed to the standard Vilnius system and so a program
to establish a set of Vilnius standard stars in the southern hemisphere
was begun.

The first results of this program are given by Forbes, Dodd \& Sullivan
(1994), hereafter Paper~II, listing a set of bright standard stars
spread across the southern sky.  To extend the standards to fainter
stars it was decided to measure the Harvard E~Region stars, which have
already been established as standards in other photometric systems
(e.g. Menzies et al. 1989).  The Harvard E Regions consist of nine
groups which contain approximately sixty stars each, ranging from $V$ =
3 to 12, centred near $ \delta \approx -45\degr $ and spaced 2 -- 3
hours apart in right ascension.  Some measurements of the Harvard
E and F~Region stars are given by {\Zdanavicius}, Dodd, Forbes \&
Sullivan (1995), hereafter Paper~III.

\section{2. OBSERVATIONS AND REDUCTIONS}

The observations were made at the Mt.\ John University Observatory
(University of Canterbury), situated near Lake Tekapo in the South
Island of New Zealand, over the period 1992 to 1993.  A set of stars
previously measured in the Vilnius system (North 1980), which are
observable from the southern hemisphere, were selected as primary
standards, to enable transformation of the local Vilnius filter system
to the standard system.  Measurements of the primary standard stars were
interspersed with those of the southern standard stars.  Certain
southern standard stars were used as extinction stars.  These were
selected from F5--G2~V spectral type stars with $ \delta \approx
-45\degr $ and measured every hour, using a variation of Nikonov's
method (Nikonov 1953, Strai\v zys 1992).  More details on the observing
and reduction procedures are given in Paper~I.

To date, \totalnumberofstars\ stars have been observed as possible
standard stars.  The following criteria were applied to select the stars
that have been measured to acceptable precision and accuracy.  First,
the star must have been measured on at least three different nights,
enabling a mean value and standard deviation to be calculated.  Next,
the standard deviation of the mean for the $V$ magnitude and colors of
the star must be less than 0.02~mag.  The only exception was the color
$U \nns - \nns P$, whose rejection threshold was relaxed to 0.025~mag.
The slightly larger errors in the $U$ filter observations imply that
longer integration times are required for measurements using that
filter.  Finally, using either the `stellar box' or $Q( \lambda )$
pseudo-spectrum methods ({\Straizys} 1992), the measurements must enable
a star to be classified within two sub-classes and within one class of
its MK spectral and luminosity classes, respectively.  This left
\numberofstars\ stars suitable for use as standards, with most of the
rejections due to an insufficient number of observations per star.  The
rejected stars remain in the observing program, to see if further
observations will enable them to meet the standard star criteria.

\section{3. CATALOG}

Table~1 lists the observations of the \numberofstars\ E region standard
stars measured in the southern Vilnius photometric system.  The first
column is the star name as given by Menzies (1989).  The next two
columns are the Epoch 2000 equatorial coordinates.  These are followed
by the Vilnius $V$ magnitude and the six color indices.  A colon
following the measurement indicates that the standard deviation of the
mean exceeds 0.015~mag; a double colon means the standard deviation is
greater than 0.02~mag.  The number of observations ($n$) of each star is
given in the next column.  The final two columns give the spectral type
as listed by Menzies et al.  (1989) and classified from the photometric
color indices by K. Zdanavi\v cius.  Close binaries were not classified
photometrically since their color indices belong to common light of the
two components.

\section{4. DISCUSSION}

The overall accuracy and the precision of the observations can be seen
in Table~2.  This shows the standard deviation $\varepsilon$ of each
color calculated using

\stdev

\noindent where $\sigma$ is the typical standard deviation of a single
observation for the southern system, $n$ is the number of observations
of a star and $N$ is the number of observed stars.  The accuracy
required for establishing a set of standard stars, i.e.\ 0.01~mag, has
clearly been achieved.
\vskip10mm

ACKNOWLEDGMENTS.  The authors thank University of Canterbury and Mt.\
John University Observatory staff for the generous use of their
facilities, the Vilnius Institute of Theoretical Physics and Astronomy
for manufacturing the filter sets and the New Zealand Lottery Board for
financing their purchase.  The New Zealand Foundation for Research,
Science and Technology and the Internal Research Grant Committee of
Victoria University of Wellington provided partial funding of this
project.  It is a pleasure to thank K. Zdanavi\v cius for photometric
classification of the observed stars and V.\ {\Straizys} for his help
and encouragement in establishing this program.

\vfil\eject

\vbox{
\Table {1}
{~Catalog of standard stars in the southern Vilnius photometric system (E~regions)}
\tabskip 33pt minus 33pt\rmn
\halign to \hsize{
#\hfil&#&#&\hfil#\hfil&#&#&#&#&#&#&\hfil#&#&#\cr
\Ruler
\HeaderTableOneE
\Ruler
E136           &01\Hr 15\Min 07\Sec &\m 45\Deg 32$'$ & 5.05    &0.49     &0.61     &0.76     &0.32     &0.20     &0.51    & 43     &G0\ts V      &F9.5\ts IV \cr
\Ruler
\BlankLineTableOne
E345\Notes~~     &06\ns 25\ns 43    &\m 48\ns 11    & 5.78\ssd &0.55\lsd &0.68     &0.23     &0.10     &0.04     &0.12    &  5     &B9\ts V      &           \cr
E347             &06\ns 49\ns 54    &\m 46\ns 37    & 5.12\ssd &0.47\lsd &0.58     &0.63     &0.29     &0.18     &0.45    &  4     &F5\ts V      &F5\ts V  \cr
\Ruler
\BlankLineTableOne
E428             &09\ns 30\ns 33    &\m 45\ns 34    & 6.60     &0.64\lsd &1.16     &1.54     &0.49     &0.39     &0.82    &  8     &K2-3\ts III  &K2\ts III\cr
E429             &09\ns 20\ns 15    &\m 45\ns 10    & 6.67     &0.66\lsd &1.21     &1.61     &0.49     &0.41     &0.84    &  8     &K2\ts III    &K2\ts III\cr
E439             &09\ns 12\ns 14    &\m 45\ns 50    & 6.66     &0.56     &0.59     &0.60     &0.28     &0.17     &0.45    & 32     &F3-5\ts V    &F3\ts V  \cr
E441\Notes~~     &09\ns 15\ns 14    &\m 45\ns 33    & 6.27     &0.49\ssd &0.61     &0.20     &0.10     &0.05     &0.15    &  8     &B8\ts V      &           \cr
E444             &09\ns 22\ns 24    &\m 46\ns 03    & 5.74     &0.57\ssd &0.89     &1.14     &0.44     &0.27     &0.67    &  8     &G6-8\ts III  &G7\ts III\cr
E445             &09\ns 29\ns 22    &\m 43\ns 11    & 6.61     &0.54\ssd &0.62     &0.66     &0.29     &0.17     &0.46    &  8     &F5\ts IV     &F6\ts IV \cr
E446             &09\ns 31\ns 19    &\m 47\ns 57    & 6.56     &0.53\ssd &0.73     &0.50     &0.20     &0.12     &0.30    &  8     &A9\ts IV-V   &A9\ts V  \cr
E447             &09\ns 38\ns 01    &\m 43\ns 11    & 5.51     &0.54\ssd &0.88     &1.25     &0.47     &0.29     &0.69    &  8     &G8\ts II     &G8.5\ts III\cr
E478             &09\ns 14\ns 08    &\m 44\ns 09    & 5.88     &0.27\ssd &0.45     &0.17     &0.08     &0.03     &0.10    &  8     &B5\ts V      &B6\ts V  \cr
E479\Notes       &09\ns 14\ns 24    &\m 43\ns 14    & 5.27     &0..25\ssd &0.39     &0.16     &0.07     &0.03     &0.10    &  8     &B3-5\ts V    &           \cr
E493             &09\ns 11\ns 33    &\m 46\ns 35    & 5.82     &0.10\ssd &0.19     &0.11     &0.05     &0.01     &0.06    &  8     &B2\ts III-IV &B1.5\ts V  \cr
E494             &09\ns 13\ns 34    &\m 47\ns 20    & 5.95     &0.55\lsd &0.73     &0.25     &0.10     &0.04     &0.14    &  6     &B9\ts V      &B9.5\ts V  \cr
E496             &09\ns 16\ns 58    &\m 45\ns 01    & 6.78     &0.57\lsd &0.81     &0.29     &0.11     &0.05     &0.13    &  6     &A0\ts V      &A0.5\ts V  \cr
E497             &09\ns 23\ns 25    &\m 48\ns 17    & 6.32     &0.27\lsd &0.35     &0.16     &0.08     &0.03     &0.11    &  6     &B3\ts IV     &B4\ts IV \cr
E498             &09\ns 23\ns 39    &\m 46\ns 55    & 6.26     &0.43\lsd &0.51     &0.19     &0.09     &0.04     &0.13    &  6     &B7\ts III    &B7\ts V  \cr
\Ruler
\BlankLineTableOne
E540             &12\ns 03\ns 38    &\m 42\ns 26    & 5.17     &0.52     &0.58     &0.60     &0.27     &0.16     &0.43    & 52     &F5\ts V      &F4\ts V  \cr
\Ruler
\BlankLineTableOne
E612             &14\ns 56\ns 14    &\m 44\ns 05    & 7.92     &0.61     &0.83     &0.45     &0.17     &0.10     &0.21    &  4     &A3\ts IV     &A5\ts V  \cr
E613             &14\ns 55\ns 25    &\m 44\ns 20    & 8.02     &0.67     &0.81     &0.49     &0.19     &0.12     &0.26    &  4     &A4\ts V      &A5.5\ts IV \cr
E614             &14\ns 53\ns 04    &\m 43\ns 44    & 8.16     &0.66     &0.81     &0.49     &0.20     &0.13     &0.29    &  4     &A5\ts V      &A6.5\ts IV \cr
E633             &14\ns 38\ns 12    &\m 45\ns 52    & 6.90     &0.64\lsd &1.18     &1.52     &0.48     &0.38     &0.82    &  4     &K2\ts III    &K2\ts III\cr
E639\Notes       &14\ns 30\ns 08    &\m 45\ns 19    & 5.51     &0.53     &0.60     &0.20     &0.10     &0.04     &0.12    & 15     &B8\ts V      &           \cr
E640             &14\ns 36\ns 18    &\m 46\ns 15    & 5.58     &0.76     &1.34     &1.93     &0.59     &0.49     &0.99    & 14     &K3\ts III    &K3.5\ts III\cr
E641             &14\ns 39\ns 11    &\m 46\ns 35    & 6.11     &0.51     &0.58     &0.69     &0.30     &0.19     &0.48    &  7     &F6\ts V      &F7\ts V  \cr
E642\Notes~~     &14\ns 45\ns 57    &\m 44\ns 52    & 6.95     &0.56     &0.79     &0.32     &0.14     &0.07     &0.15    &  5     &A0\ts V      &A0\ts V  \cr
E643             &14\ns 46\ns 28    &\m 47\ns 26    & 5.77     &0.61     &0.83     &0.36     &0.14     &0.06     &0.16    &  7     &A1\ts V      &A1.5\ts V  \cr
E644             &14\ns 47\ns 31    &\m 43\ns 33    & 6.34     &0.63     &1.00     &1.32     &0.49     &0.32     &0.76    & 17     &G8\ts III    &G9.5\ts III\cr
\Ruler
}}
\vfill
\continued{1}
\vbox{
\tabskip 33pt minus 33pt\rmn
\halign to \hsize{
#\hfil&#&#&\hfil#\hfil&#&#&#&#&#&#&\hfil#&#&#\cr
\Ruler
\HeaderTableOneE
\Ruler
E645          &14\Hr 56\Min 24\Sec    &\m 44\Deg 42$'$    & 6.77     &0.62     &0.83     &0.35     &0.13     &0.07 &0.15  & 35     &A1\ts V      &A1.5\ts V  \cr
E646\Notes       &14\ns 59\ns 18    &\m 43\ns 28    & 7.08     &0.56     &0.66     &0.57     &0.24     &0.16     &0.38    &  4     &F2\ts III    &           \cr
E647             &14\ns 59\ns 45    &\m 43\ns 49    & 6.60     &0.49     &0.55     &0.66     &0.29     &0.18     &0.46    & 65     &F7\ts IV-V   &F5\ts V  \cr
E679\Notes       &14\ns 26\ns 10    &\m 45\ns 23    & 4.38     &0.72     &0.81     &0.57     &0.27     &0.17     &0.46    &  9     &F7\ts        &           \cr
E680             &14\ns 27\ns 12    &\m 46\ns 08    & 5.85     &0.62     &0.76     &0.58     &0.21     &0.14     &0.32    &  9     &A1m          &A6\ts V    \cr
E681\Notes       &14\ns 28\ns 51    &\m 47\ns 59    & 6.41     &0.43     &0.53     &0.19     &0.08     &0.05     &0.11    &  8     &ApSi         &           \cr
E682             &14\ns 35\ns 09    &\m 46\ns 28    & 6.92     &0.68     &0.68     &0.23     &0.13     &0.06     &0.16    &  8     &B9\ts V      &B9\ts III  \cr
E684             &14\ns 35\ns 49    &\m 46\ns 49    & 6.82     &0.61     &0.89     &1.18     &0.46     &0.29     &0.71    &  8     &G8\ts III    &G7\ts III\cr
E687             &14\ns 37\ns 54    &\m 48\ns 01    & 6.66     &0.53     &0.58     &0.22     &0.13     &0.06     &0.17    &  4     &B8\ts V      &B8\ts III-IV \cr
E688\Notes       &14\ns 40\ns 18    &\m 45\ns 48    & 6.64     &0.42\ssd &0.51     &0.20     &0.04     &0.07     &0.15    &  3     &ApSi\ts      &           \cr
E690             &14\ns 50\ns 58    &\m 42\ns 49    & 6.84     &0.65     &0.84     &0.43     &0.15     &0.09     &0.21    &  4     &A2\ts V      &A4\ts V  \cr
E691\Notes       &14\ns 51\ns 37    &\m 43\ns 34    & 4.34     &0.28     &0.33     &0.15     &0.07     &0.04     &0.10    &  8     &B5\ts IV     &           \cr
E695             &14\ns 59\ns 26    &\m 43\ns 10    & 6.14     &0.74     &0.72     &0.78     &0.32     &0.22     &0.54    &  6     &F7\ts II     &F7\ts II \cr
E6100             &14\ns 54\ns 00    &\m 43\ns 36    & 8.23     &0.59\ssd &0.91     &1.30     &0.48     &0.31     &0.73   &  4     &G8\ts II     &G9\ts III\cr
\Ruler
\BlankLineTableOne
E748             &17\ns 24\ns 43    &\m 45\ns 01    & 6.70     &0.54     &0.60     &0.58     &0.26     &0.15     &0.40    & 95     &F3\ts V      &F2.5\ts V  \cr
E765             &17\ns 37\ns 53    &\m 42\ns 34    & 7.22     &0.48     &0.65     &0.83     &0.34     &0.22     &0.56    & 55     &G2-3\ts V    &G2\ts IV-V \cr
\Ruler
\BlankLineTableOne
E869             &19\ns 55\ns 15    &\m 41\ns 52    & 4.14\ssd &0.62     &1.00     &1.35     &0.47     &0.34     &0.76    &  4     &K0\ts II-III &K0\ts III\cr
\Ruler
\BlankLineTableOne
E924             &22\ns 32\ns 11    &\m 43\ns 16    & 6.99     &0.59     &0.95     &1.23     &0.45     &0.29     &0.72    &  6     &K0\ts III    &G9.5\ts III \cr
E929             &22\ns 33\ns 54    &\m 44\ns 41    & 6.87     &0.58     &0.87     &1.16     &0.45     &0.28     &0.71    &  7     &G8\ts III    &G6\ts III\cr
E930             &22\ns 53\ns 15    &\m 45\ns 09    & 6.93     &0.63     &1.11     &1..42     &0.49     &0.34     &0.78    &  7     &K0\ts III    &K0.5\ts III\cr
E940             &22\ns 30\ns 43    &\m 44\ns 06    & 7.00     &0.62     &0.84     &0.37     &0.14     &0.07     &0.15    &  5     &A0-1\ts V    &A2\ts V  \cr
E942             &22\ns 36\ns 41    &\m 43\ns 28    & 6.81     &0.52     &0.57     &0.70     &0.31     &0.20     &0.49    & 37     &F8\ts V      &F6\ts V  \cr
E943             &22\ns 42\ns 43    &\m 44\ns 15    & 6.15     &0.58     &0.90     &1.19     &0.45     &0.29     &0.71    &  7     &K0\ts III    &G8\ts III\cr
E944             &22\ns 45\ns 41    &\m 46\ns 33    & 5.60     &0.69     &1.23     &1.71     &0.55     &0.41     &0.89    &  7     &K2\ts III    &K2.5\ts III\cr
E948             &22\ns 30\ns 15    &\m 42\ns 17    & 6.99     &0.57     &0.74     &0.25     &0.11     &0.04     &0.11    &  7     &B9.5\ts V    &B9.5\ts V  \cr
E949             &22\ns 47\ns 09    &\m 41\ns 42    & 6.92     &0.67     &1.22     &1.66     &0.53     &0.41     &0.88    &  6     &K2\ts III    &K2.5\ts III\cr
E963\Notes       &22\ns 42\ns 37    &\m 47\ns 12    & 6.05     &0.49     &0.59     &0.74     &0.32     &0.20     &0.53    &  4     &G0\ts V      &           \cr
\Ruler
}}
\vfill

\continued{1}
\vbox{
\tabskip 33pt minus 33pt\rmn
\halign to \hsize{
#\hfil&#&#&\hfil#\hfil&#&#&#&#&#&#&\hfil#&#&#\cr
\Ruler
\HeaderTableOneE
\Ruler
E971        &22\Deg 15\Min 35\Sec    &\m 44\Deg 27$'$    & 6.18     &0.62     &0.94     &1.26     &0.48     &0.30 &0.73   &  6     &G8-K0\ts III &G8.5\ts III \cr
E972             &22\ns 23\ns 07    &\m 45\ns 56    & 5.69     &0.56     &0.63     &0.57     &0.25     &0.15     &0.40    &  6     &F0\ts V      &F1.5\ts V  \cr
E973             &22\ns 23\ns 25    &\m 46\ns 40    & 6.96     &0.65     &0.83     &0.36     &0.13     &0.08     &0.16    &  4     &A1\ts V      &A1.5\ts V  \cr
E979             &22\ns 43\ns 30    &\m 41\ns 25    & 4.91     &0.61     &0.97     &1.26     &0.47     &0.31     &0.74    &  5     &K0\ts III    &G9.5\ts III\cr
\Ruler
}}

\vbox{
\vskip2truemm
\noindent
{\tabfont N~o~t~e~s~ (from Menzies 1989):}\hfil\break
\noindent
{\rmn E345 -- double, 6.0 + 8.3, 1.5\sarcsec\hfil\break
E441 -- double, 6.6 + 7.7, 1\sarcsec\hfil\break
E479 -- double, 5.2 + 9.6, 6\sarcsec\hfil\break
E639 -- companion approx 8\sarcsec\ NW\hfil\break
E642 -- double, 7.3 + 12.0, 35\sarcsec, companion not
included\hfil\break E646 -- double, 7.2 + 11.2, 4\sarcsec\hfil\break
E679 -- double, 5.2 + 5.3, 0.3\sarcsec, cpm\hfil\break
E681 -- companion 8\sarcsec\ NW\hfil\break
E688 -- companion approx 8\sarcsec\ W\hfil\break
E691 -- double, 0.1\sarcsec\hfil\break
E963 -- double, 6.4 + 9.8, 8\sarcsec, cpm\hfil\break
}}

\Table {2}{~Summary of precision of the Vilnius southern standard
stars.}
\vbox{\tabskip 33pt minus 33pt\rmn
\halign to \hsize{
 \hfil# \hfil&# &# &# &# &# &# \cr
\Ruler
\HeaderTableTwo
\Ruler
 0.008  & 0.008 & 0.005 & 0.004 & 0.004 & 0.004 & 0.005 \cr
\Ruler
}}

\References
\ref
Forbes M.\ts C., Dodd R.\ts J., Sullivan D.\ts J. 1993,
Baltic Astronomy, 2, 246
\ref
Forbes M.\ts C., Dodd R.\ts J., Sullivan D.\ts J. 1994,
Baltic Astronomy, 3, 227
\ref
Zdanavi\v cius K., Dodd R.\ts J., Forbes M.\ts C., Sullivan D.\ts J.
1995, Baltic Astronomy, 4, 25
\ref
Menzies J.\ts W., Cousins A.\ts W.\ts J., Banfield R.\ts M., Laing J.\ts D. 1989,
Circ. South African Astron. Obs., No.~13, 1
\ref
Nikonov V.\ts B. 1953, Bull. Abastumani Obs., No.~14
\ref
North P. 1980, A\&AS, 41, 395
\ref
Strai\v zys V. 1992, Multicolor Stellar Photometry,
Pachart Publishing House, Tucson, Arizona
\bye